\documentclass[twocolumn,showpacs,preprintnumbers,amsmath,amssymb]{revtex4}
\usepackage{color}
\usepackage{graphics}% Include figure files
\usepackage{graphicx}% Include figure files
\usepackage{dcolumn}% Align table columns on decimal point
\usepackage{bm}% bold math

\renewcommand\textfraction 0
\renewcommand\topfraction 1
\renewcommand\bottomfraction 1
\begin{document}

\title{Structural and physical properties of SrMn$_{1-x}$Ru$_{x}$O$_{3}$ perovskites}
 \author{S. Kolesnik, B. Dabrowski, and O. Chmaissem}
 \affiliation{Department of Physics, Northern Illinois University, DeKalb, IL 60115}
 \affiliation{Materials Science Division, Argonne National Laboratory, Argonne, IL 60439}

\date{\today}
\begin{abstract}
We combine the results of magnetic and transport measurements with neutron diffraction data to construct the structural and magnetic phase diagram of
the entire family of SrMn$_{1-x}$Ru$_{x}$O$_3$ ($0 \leqslant x \leqslant 1$) perovskites. We have found antiferromagnetic ordering of the C type for
lightly Ru-substituted materials ($0.06 \leqslant x \leqslant 0.5$) in a similar manner to $R_{y}$Sr$_{1-y}$MnO$_3$ ($R$=La, Pr), due to the
generation of Mn$^{3+}$ in both families of manganite perovskites by either $B$-site substitution of Ru$^{5+}$ for Mn$^{4+}$ or $A$-site substitution
of $R^{3+}$ for Sr$^{2+}$. This similarity is driven by the same ratio of $d^4$/$d^3$ ions in both classes of materials for equivalent substitution
level. In both cases, a tetragonal lattice distortion is observed, which for some compositions ($0.06 \leqslant x \leqslant 0.2$) is coupled to a
C-type AF transition and results in a first order magnetic and resistive transition. Heavily substituted SrMn$_{1-x}$Ru$_{x}$O$_3$ materials are
ferromagnetic due to dominating exchange interactions between the Ru$^{4+}$ ions. Intermediate substitution ($0.6 \leqslant x \leqslant 0.7$) leads
to a spin-glass behavior instead of a quantum critical point reported previously in single crystals, due to enhanced disorder.
 \end{abstract}

\pacs{75.30.Kz, 75.50.Ee, 75.50.Lk, 81.30.Dz}

%}
 \maketitle

\section{INTRODUCTION}

The substitution of Ru in perovskite manganites has been demonstrated to lead to a variety of interesting physical phenomena. For colossal
magnetoresistance manganites  La$_{0.5}$Sr$_{0.5}$Mn$_{1-y}$Ru$_{y}$O$_{3}$ with ferromagnetic matrix and
La$_{0.45}$Sr$_{0.55}$Mn$_{1-y}$Ru$_{y}$O$_{3}$ with antiferromagnetic matrix, the low Ru doping  $0.05 \leqslant y \leqslant 0.15$ induces an
enhanced ferromagnetism with an increasing Curie temperature $T_C$.\cite{Ying06} Ru ions in these materials exist mainly in the form of Ru$^{4+}$
with a small quantity of Ru$^{5+}$. A ferromagnetic  exchange interaction between Mn$^{3+}$ and Ru$^{4+}$(Ru$^{5+}$) has been attributed to this
enhancement of ferromagnetism.\cite{Ying06} Similarly in charge ordered Nd$_{0.5}$Sr$_{0.5}$MnO$_{3}$ $T_C$ is significantly increased  by
substitution of Ru$^{4+}$, but the charge ordering can be destroyed.\cite{Vanitha99} The incorporation of Ru ions in CaMn$_{1-y}$Ru$_{y}$O$_3$
perovskites can induce ferromagnetism in a large substitution range $0.1 \leqslant y \leqslant 0.8$ with a maximum Curie temperature $T_C=210$ K for
$y=0.4$ and a metallic character for $0.2 \leqslant y \leqslant 0.4$.\cite{Maignan01} By substitution of only a few percent of Ru with Mn in
Sr$_{3}($Ru$_{1-y}$Mn$_{y})_{2}$O$_{7}$ the ground state can be switched from a paramagnetic metal to an antiferromagnetic insulator.\cite{Mathieu05}

The study of SrRu$_{1-y}$Mn$_{y}$O$_3$ single crystals in the limited (Ru-rich) range of compositions $0 \leqslant y \leqslant 0.6$ has shown that
the Mn substitution can drive the system from the itinerant ferromagnetic (F) state for SrRuO$_3$ through a ``quantum critical point'' at $y_c=0.39$
to an insulating antiferromagnetic (AF) state.\cite{Cao05} Sahu {\em et al.}\cite{Sahu02jap,Sahu02prb} reported a contradictory finding that the
ferromagnetic state may still be observed with higher Mn contents including SrRu$_{0.5}$Mn$_{0.5}$O$_3$ for polycrystalline samples prepared in air
at 1200$^{\circ}$C. A more complicated phase diagram with the coexistence of F and AF phases in a wide range of substitution and a large
magnetoresistance have been reported by Zhang {\em et al.} \cite{Zhang07} for polycrystalline samples prepared in air at 1150$^{\circ}$C. The
discrepancy can be traced to the highly inhomogeneous polycrystalline samples obtained to date, containing a large amount of SrRuO$_3$, for which
only a fraction of the FM phase changes but not the magnetic phase or transition temperature (see: Ref. 6, Fig. 1; Ref. 7, Fig. 2, Ref. 8, Fig. 3,
and Ref. 9, Fig. 3).

The end member of the SrMn$_{1-x}$Ru$_{x}$O$_3$ family, SrRuO$_3$ is a unique ferromagnetic metal among 4$d$ transition metal based perovskite
oxides. Most dopants for low spin Ru$^{4+}$ ($t_{2g}^4$) decrease the ferromagnetic Curie temperature from 163 K, except for
Cr.\cite{Pi02,Dabrowski05,Williams06} The other end member, a cubic perovskite SrMnO$_3$ is a G-type antiferromagnet with $T_N = 233$~K. The
oxidation state of Mn in the latter material is also 4+. When this valency of Mn is preserved (e.g. as in Sr$_{1-y}$Ca$_{y}$MnO$_3$), then the G-type
AFM ordering is observed in the cubic, tetragonal and orthorhombic crystal structures.\cite{Chmaissem01} $T_N$ is suppressed by the deviation of the
Mn-O-Mn bond angle from 180$^{\circ}$ and by the variance of the average size of the $A$-site ion via changes in the Sr/Ca
ratio.\cite{Chmaissem01,Dabrowski03}

The substitution of Ru$^{5+}$ for Mn$^{4+}$ in SrMnO$_3$ was considered \cite{Maignan02} to stabilize the cubic perovskite structure by the induced
Mn valency shift, corresponding to electron doping by Mn$^{3+}$ in the Mn$^{4+}$ matrix. The $L_{2,3}$-edge absorption spectroscopy of Ru and Mn in
Ru-rich SrRu$_{1-y}$Mn$_{y}$O$_3$ ($0\leqslant y \leqslant 0.5$) has revealed the mixed-valence of both Mn$^{3+}$/Mn$^{4+}$ and
Ru$^{4+}$/Ru$^{5+}$.\cite{Sahu02prb} $^{55}$Mn NMR on SrRu$_{0.9}$Mn$_{0.1}$O$_3$ has demonstrated that Mn exists in an intermediate Mn$^{3+/4+}$
valence state due to fast electron hopping.\cite{Han06}

In this study, we investigate the complete solubility range of polycrystalline SrMn$_{1-x}$Ru$_{x}$O$_3$ samples and construct the phase diagram of
structural, magnetic, and conducting properties. The polycrystalline samples were characterized by neutron diffraction, magnetic, transport and
thermoelectric experiments. The incorporation of Ru in the SrMnO$_3$ matrix ($0.06 \leqslant x \leqslant 0.2$) results in a phase transition to a
C-type antiferromagnetic state accompanied by a cubic-tetragonal transition. At slightly higher substitutions ($0.3 \leqslant x \leqslant 0.5$) the
structural transition temperature is higher than the AF transition temperature. The intermediate substitution level ($0.6 \leqslant x \leqslant 0.7$)
induces a spin-glass behavior, due to competing ferro- and antiferromagnetic interactions in the tetragonal structure. Close to the maximum Ru
substitution ($0.8 \leqslant x \leqslant 1$) the material becomes ferromagnetic in the orthorhombic structure.

\section{EXPERIMENTAL DETAILS}
\begin{figure}
 \resizebox{8.5cm}{!}{\includegraphics{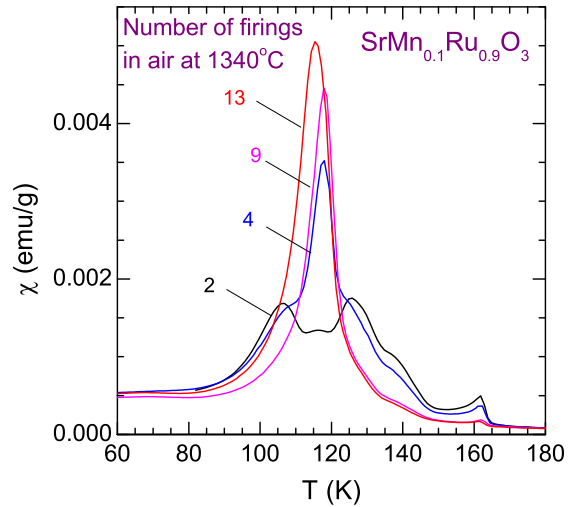}}
\caption{\label{ac} (Color online) Magnetic ac susceptibility for the SrMn$_{0.1}$Ru$_{0.9}$O$_3$ sample after increasing number of firings. A
magnetically single-phase material is obtained after 13 grindings and firings. }
\end{figure}

Samples with $x \leqslant 0.5$ were prepared using a two-step synthesis method developed for similar kinetically stable perovskites.\cite{Hinks88}
First, oxygen-deficient samples were prepared in argon at $T = 1300 - 1400^{\circ}$C. The samples were then annealed in air at lower temperatures to
achieve stoichiometric compositions with respect to the oxygen content. The samples  with $x > 0.5$ were prepared in air at $1330-1340^{\circ}$C with
many (up to 14) intermittent grindings due to difficulty of achieving homogeneous material. An excess of RuO$_2$ was added to compensate for Ru loss
due to sublimation at these high temperatures. The process of formation of single-phase and homogeneous material was monitored with x-ray diffraction
(Rigaku D/MAX diffractometer) and ac susceptibility (Physical Property Measurement System Model 6000, Quantum Design) measurements. After a few
firings, x-ray diffraction indicated formation of single-phase material, though the ac susceptibility measurements clearly showed peaks related to
multiple magnetic transitions and hence highly inhomogeneous samples. Fig.~\ref{ac} shows a sequence of ac susceptibility measurements for the
SrMn$_{0.1}$Ru$_{0.9}$O$_3$ sample, which demonstrates the gradual improvement of sample quality. This difficulty to achieve research quality
polycrystalline materials may explain the discrepancy in the properties of single crystals and bulk samples reported to date.

 The ac susceptibility, resistivity, thermal conductivity and Seebeck coefficient were measured using a Physical Property Measurement
System Model 6000 (Quantum Design). The dc magnetization was measured using a Magnetic Property Measurement System Model MPMS-7 (Quantum Design).
Time-of-flight neutron powder diffraction data were collected at 300 K (room temperature) for all members of the SrMn$_{1-x}$Ru$_{x}$O$_3$ series on
the Special Environment Powder Diffractometer (SEPD)\cite{sepd} at the Intense Pulsed Neutron Source (IPNS) at Argonne National Laboratory.  Data
were collected, for the $x = 0.2$, 0.7 and 0.9 samples, at several temperatures between 10 and 320 K using a closed cycle refrigerator.  In the
refinements, high-resolution backscattering data, from 0.5 to 4~\AA~$d$-spacing were analyzed using the Rietveld method and the General Structure
Analysis System (GSAS) code.\cite{gsas}  Absorption, background and peak width parameters were refined, together with the lattice parameters, atomic
positions, and isotropic and anisotropic temperature factors for the cations and oxygen atoms, respectively.

\begin{figure}
 \resizebox{8.5cm}{!} {\includegraphics{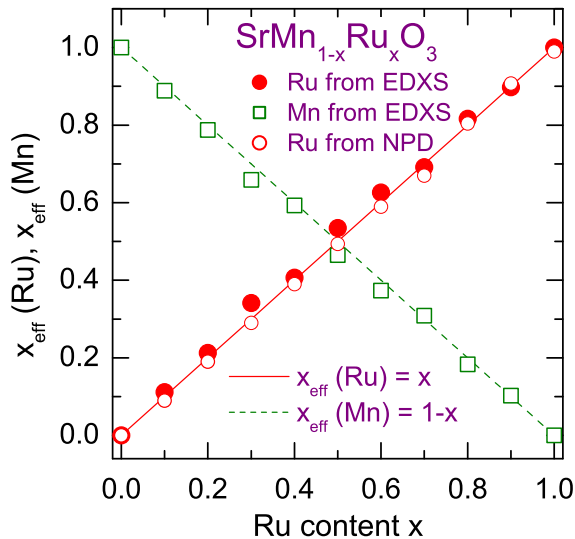} }
\caption{\label{eds} (Color online) The effective contents of Ru and Mn ions for SrMn$_{1-x}$Ru$_{x}$O$_3$ from the energy dispersive x-ray spectra
(EDXS). The straight lines are the nominal contents of both cations. The statistical errors are within the marker symbols. The refined Ru contents
from neutron powder diffraction (NPD) are shown as open circles.}
\end{figure}

The cationic ratio was determined by energy dispersive x-ray spectroscopy (EDXS) analysis in a Hitachi S-4700-II scanning electron microscope at the
Electron Microscopy Center, Argonne National Laboratory. Typically, 5 spot spectra were collected across the surface of sintered pellets.
Fig.~\ref{eds} presents the effective contents $x_{\rm eff}$ of Ru and Mn ions calculated from the EDXS spectra using a normalization condition
$x_{\rm eff}$(Ru)+$x_{\rm eff}$(Mn)=1. We observe a good agreement with the nominal compositions drawn as straight lines in Fig.~\ref{eds}. This
result is strongly supported by the refined Ru occupancies from the neutron powder diffraction data.

\section{Results and Discussion}

\subsection{Neutron powder diffraction and structural details}

\begin{figure*}
 \resizebox{15cm}{!} {\includegraphics{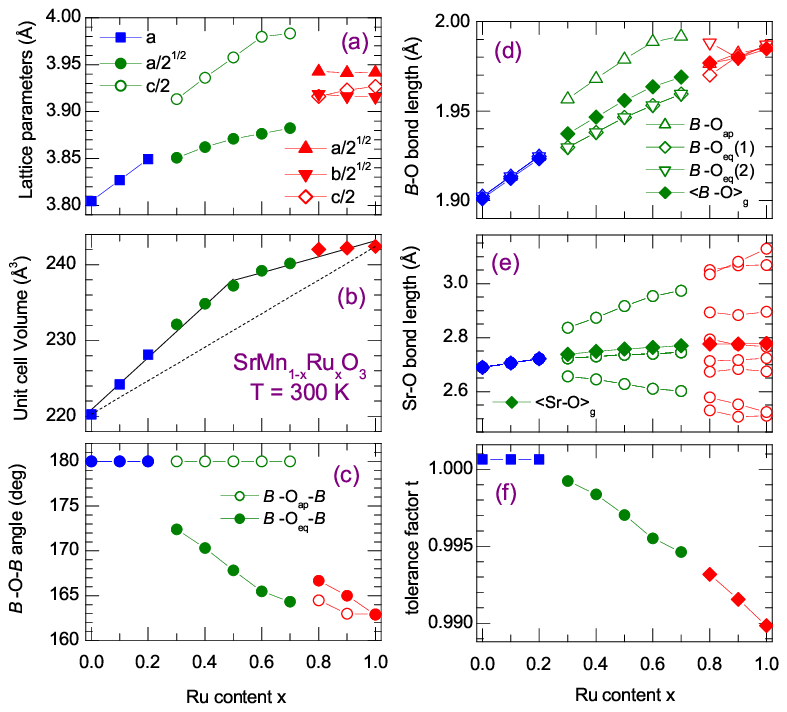} }
\caption{\label{latt} (Color online) Room temperature structural parameters refined from neutron diffraction for SrMn$_{1-x}$Ru$_{x}$O$_3$. $B$ is
the $B$-site ion Ru or Mn. (a,b,f) Squares denote a cubic $Pm-3m$ structure, circles - tetragonal $I4/mcm$, diamonds - orthorhombic $Pbnm$. (d,e)
Open symbols: individual bond lengths, full symbols: geometrical average. (f) The tolerance factor is defined as $<$Sr-O$>$/($<B$-O$>\sqrt{2}$). The
dashed line in (b) represents Vegard's law, the solid lines in (b) are linear fits to the data (see: text). }
\end{figure*}
The well-known perovskite structures of the two end-members of the series, namely SrMnO$_{3}$ and SrRuO$_{3}$, have frequently been described as
crystallizing in the cubic   and orthorhombic   space group symmetries, respectively.  The Ru spins, in SrRuO$_{3}$, do not localize (are itinerant)
and the material is viewed as an itinerant ferromagnet below 163 K.  On the other hand, in SrMnO$_{3}$, the Mn spins become localized below 233 K in
a G-type antiferromagnetic structure.

Room temperature structural refinements (T = 300 K) for all members of the SrMn$_{1-x}$Ru$_{x}$O$_3$ series demonstrate that, upon increasing $x$,
the symmetry changes from cubic $Pm-3m$ ($x \leqslant 0.2$) to tetragonal $I4/mcm$ (for $0.3 \leqslant x \leqslant 0.7$) to orthorhombic $Pbnm$ for
($x \geqslant 0.8$) in a good agreement with the different magnetic and resistive properties of the materials discussed in the next subsections.
Refined Ru and Mn site occupancies were in agreement with the nominal values, within 1-3 standard deviations, as shown in Fig.~\ref{eds}. Refined
structural parameters are presented in Fig.~\ref{latt} as a function of composition. The lattice parameters [Fig.~\ref{latt}(a)] display overall
increase associated with a larger ionic size of Ru$^{4+}$ (the average bond length $<$Ru$^{4+}$-O$>$=1.985~\AA) than Mn$^{4+}$ (the average bond
length $<$Mn$^{4+}$-O$>$=1.903~\AA) as observed in Fig.~\ref{latt}(d). The sequence of structural transitions from high symmetry cubic $Pm-3m$ to low
symmetry orthorhombic $Pbnm$ is thus a consequence of decreasing tolerance factor of the perovskite structure $t(x)=<$Sr-O$>/ \sqrt{2}<B$-O$>$
($B$=Mn,Ru) from 1 to 0.99 [Fig.~\ref{latt}(f)]. Similar sequence of transitions was observed for Sr$_{1-x}$Ca$_x$MnO$_3$, for which decrease of
tolerance factor was a result of smaller ionic size of Ca than Sr.\cite{Chmaissem01} Since neutron diffraction found no evidence for Mn/Ru cation
ordering at any $x$, the volume would be expected to vary linearly with $x$ according to the Vegard's law, presented as a dashed line in
Fig.~\ref{latt}(b). However, the unit cell volume exhibits deviations from the linear behavior especially when crossing from the Mn-rich side to the
Ru-rich side of the phase diagram [Fig.~\ref{latt}(b)]. These deviations can be solely explained by geometrical considerations of the charge transfer
Ru$^{4+}$ (0.62~\AA) + Mn$^{4+}$(0.53~\AA) $\rightarrow$ Ru$^{5+}$ (0.565~\AA) + Mn$^{3+}$(0.645~\AA) from the fact that the average ionic size of a
Ru$^{5+}$,Mn$^{3+}$ pair (0.605~\AA) is larger than that of a Ru$^{4+}$,Mn$^{4+}$ pair (0.575~\AA). Following the procedure developed by Williams
{\em et al.} \cite{Williams06} for SrRu$_{1-y}$Cr$_{y}$O$_3$, we made linear fits to the data in Fig.~\ref{latt}(b) and obtained a good agreement of
the charge transfer model with the data. A similar phenomenon has recently been observed \cite{Taniguchi08} in CaRu$_{1-y}$Mn$_{y}$O$_3$ and also
interpreted in terms of mixed valence Ru$^{4+}$, Ru$^{5+}$, Mn$^{4+}$, and Mn$^{3+}$ ions. The latter compound preserves its orthorhombic $Pnma$
structure within the entire composition range.

Another anomalous feature observed in the data is a large increase of the average $<$Sr-O$>$ bond length from 2.69 to 2.78~\AA~ [Fig.~\ref{latt}(e)].
In order to interpret the ``abnormal'' behavior of the $<$Sr-O$>$ bond length, we performed simple Bond Valence Sum calculations \cite{Brown85} from
which we find the calculated oxidation state $v$(Sr) of Sr to decrease from 2.5 to 2.0 as the Ru content increases from 0 to 1.  The unphysical
values of the Sr oxidation state (i.e., when $v$(Sr)$>$2) may be interpreted as evidence for the presence of significant strains in the Mn-rich side
of the phase diagram with the strains relaxing as a function of increased Ru content. Strain relaxation would then occur through a series of
structural distortions from heavily stressed cubic to moderately stressed tetragonal and finally to ``stress-free'' orthorhombic structures. Further
evidence for stress relaxation may be observed in the behavior of the unit cell volume as seen in the change of slope in Fig.~\ref{latt}(b).
Additionally, a changeable $<$Sr-O$>$ bond length could also be due to the decreasing size of the oxygen ion as a function of hole transfer to it
from Mn, i.e., formation of the ligand holes for SrMnO$_3$ compound. The Mn to O charge transfer would not lead to a large change of the $<$Mn-O$>$
bond length as the overall amount of charge remain constant on the electronically relevant Mn-O network. Another possibility would be that simply the
$<$Sr-O$>$ bond lengths are changeable depending on the $B$-site ion of the perovskite structure. To unambiguously differentiate between these
possibilities extensive x-ray absorption spectroscopy studies would be necessary for both transition metal and oxygen ions. We point out here,
however, that peculiar magnetic properties of SrMnO$_3$ and lightly substituted compounds, which are discussed in the following sections, may be
caused by the charge transfer from Mn to O ions.

\begin{figure*}
 \resizebox{15cm}{!} {\includegraphics{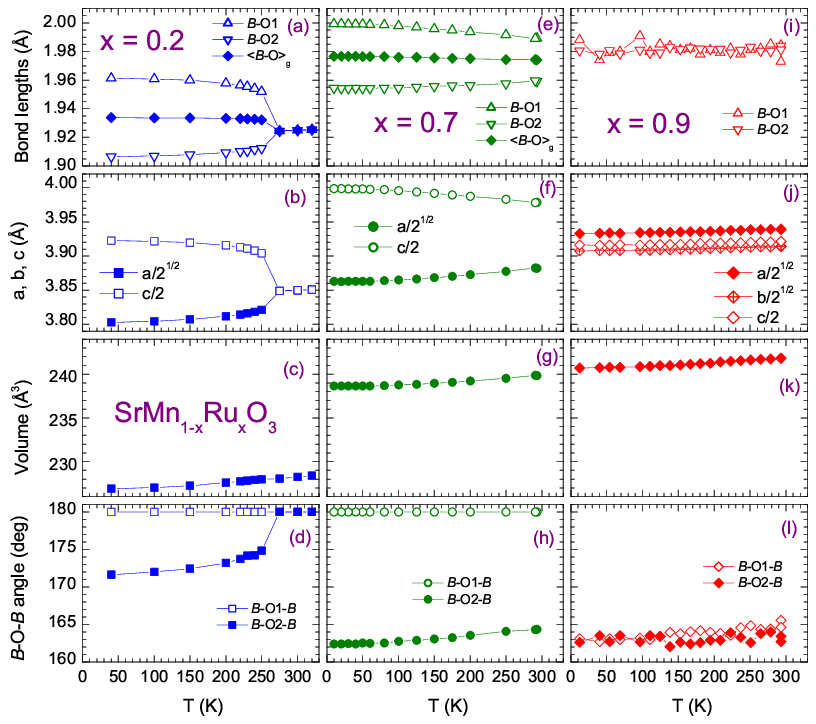} }
\caption{\label{vst} (Color online) Temperature dependence of the refined structural parameters for three SrMn$_{1-x}$Ru$_{x}$O$_3$ samples: $x=0.2$
(a-d),  $x=0.7$ (e-h), and  $x=0.9$ (i-l). $B$ is the $B$-site ion Ru or Mn. (a,e): open symbols: individual bond lengths, full symbols: geometrical
average.}
\end{figure*}

Evolution of the structure of the $x=0.2$ sample as a function of temperature is shown in Fig.~\ref{vst}.  At temperatures above 260 K, the
paramagnetic material is best described using the cubic $Pm-3m$ symmetry.  Below 260 K, a structural phase transition takes place to lower tetragonal
$I4/mcm$ space group symmetry and additional antiferromagnetic peaks become visible.  In this space group, a good fit to the magnetic intensities
could only be achieved by further lowering the magnetic symmetry to $I_p4/mc'm'$.  In this magnetic symmetry, long-range ordering of c-axis oriented
Ru/Mn spins takes place to form C-type antiferromagnetically coupled FM chains.

Temperature-dependent neutron diffraction patterns for the tetragonal $x=0.7$ and orthorhombic $x=0.9$ samples show no structural change and no extra
magnetic intensities at any temperature between 10 and 300 K in good agreement with the materials' spin-glass and itinerant ferromagnetic properties,
respectively. A decrease of the difference between individual $B$-O bonds [Fig.~\ref{vst}(e)] with increasing temperature as well as a similar effect
for lattice parameters [Fig.~\ref{vst}f)] and $B$-O-$B$ bond angles [Fig.~\ref{vst}(h)] indicate an incipient transition to the cubic phase for the
$x$ = 0.7 composition. No such behavior is observed for the orthorhombic $x$ = 0.9 composition, for which structural transitions to tetragonal and
cubic phases appear to remain at high temperatures similar to SrRuO$_3$.\cite{Kennedy98} In addition, the $x$ = 0.9 composition does not exhibit a
distinctive invar effect, which was observed below Curie temperature for SrRuO$_3$.\cite{Dabrowski05b} Suppression of the invar effect with a small
amount of Mn substitution in SrRuO$_3$ is similar to both Cr- and La-substitution,\cite{Klein06, Pietosa07} and introduction of Ru-vacancy.
\cite{Dabrowski04}

\subsection{Magnetic properties}

\begin{figure}
 \resizebox{8.5cm}{!}{\includegraphics{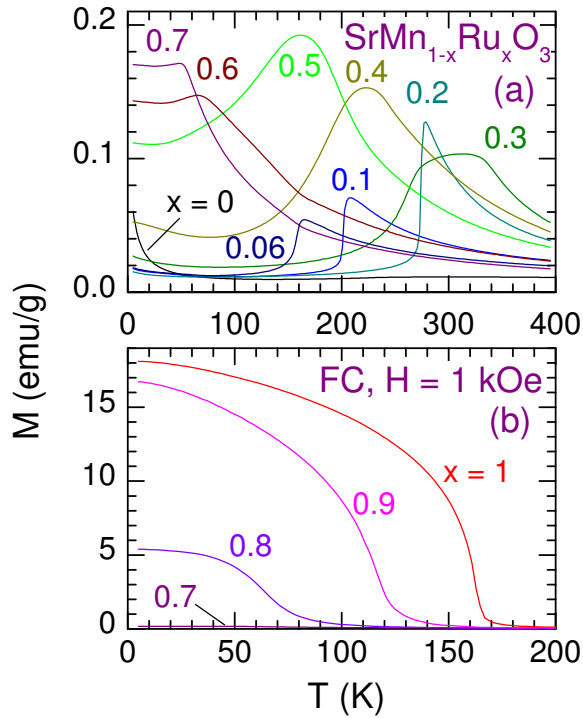}}
 \caption{\label{dc} (Color online) dc magnetization for SrMn$_{1-x}$Ru$_{x}$O$_3$ samples. }
\end{figure}

The dc magnetization measured on cooling in a magnetic field of 1 kOe is presented in Fig.~\ref{dc}. From these results we have determined N\'eel and
Curie temperatures $T_N$ and $T_C$ (defined as the temperatures for which the slope of magnetization $dM/dT$ is maximum and minimum, respectively).
On substitution of small amount of Ru for Mn ($0.06 \leqslant x \leqslant 0.2$), we observe sharp magnetic transitions from paramagnetic to a C-type
AF ordered state. The N\'eel transitions in this substitution range are coupled with the structural cubic-tetragonal transitions, i.e. $T_N = T_s$.
This type of behavior has also been observed for Sr-rich  $R_{y}$Sr$_{1-y}$MnO$_3$ ($R$=La, Pr)\cite{Chmaissem03}. For larger $x$, the structural
transition takes place at temperatures $T_s$ higher than $T_N$. As a result, the magnetic transition is not as sharp and an anomalous magnetization
is observed for $x=0.3$ in $T_N \leqslant T \leqslant T_s$ before the material becomes paramagnetic above $T_s$. Further Ru doping decreases $T_N$,
which is maximum for $x \sim 0.2-0.3$. More substitution of Ru leads to a change of the magnetic ordering from AF to F, although this boundary is not
as sharp as reported for single crystals,\cite{Cao05} but it is spread over a range of compositions ($0.6 \leqslant x \leqslant 0.7$) where a
spin-glass behavior can be observed. In this range of substitution we also observed a cusp in ac susceptibility, which supports the spin-glass
behavior.

\begin{figure}
 \resizebox{8.5cm}{!}{\includegraphics{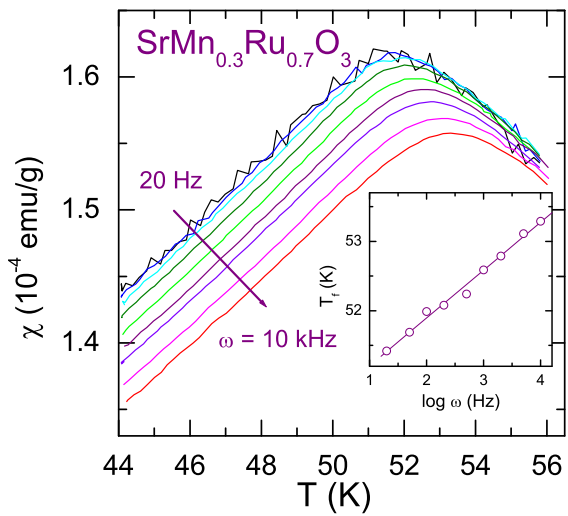}}
\caption{\label{freq} (Color online) Temperature dependence of ac susceptibility  for SrMn$_{0.3}$Ru$_{0.7}$O$_3$ at several frequencies. Inset shows
the linear dependence of $T_f$ on log frequency.}
\end{figure}

 In Fig.~\ref{freq},
we present the ac susceptibility for  SrMn$_{0.3}$Ru$_{0.7}$O$_3$ measured at several frequencies $\omega$ in an ac magnetic field of 14 Oe. One can
observe a cusp in the ac susceptibility related to spin-glass behavior, a decrease of the ac susceptibility below $T_f$ with increasing frequency,
and a shift of $T_f$ towards higher temperatures. The linear fit to $T_f(\log\omega)$ gives relative temperature shift vs. frequency $\Delta T_f/[T_f
\Delta(\log \omega)] =  0.0136 \pm 0.005$. This value is similar to those observed for the SrMn$_{1-x}$Fe$_{x}$O$_3$ perovskite having mixed F and AF
interactions. \cite{Kolesnik03} The spin-glass related irreversibility between the ``zero field cooled'' and ``field cooled'' magnetization can also
be observed (not shown). For lower Ru contents, ($0.4 \leqslant x \leqslant 0.6$) this kind of irreversibility, resembling a spin glass behavior, can
also be observed at temperatures below $\sim$60 K in the AF state. This points to a frustrated/disordered AF state,\cite{Kolesnik03} which is
sometimes confused with spin glass behavior.\cite{Zhang07} A closer inspection of the remanent magnetization after ``field cooling'' shows that a
slight irreversibility persists up to $T_N$ in these compositions, which points to a certain level of disorder in the AF state.

\begin{figure}
 \resizebox{8.5cm}{!} {\includegraphics{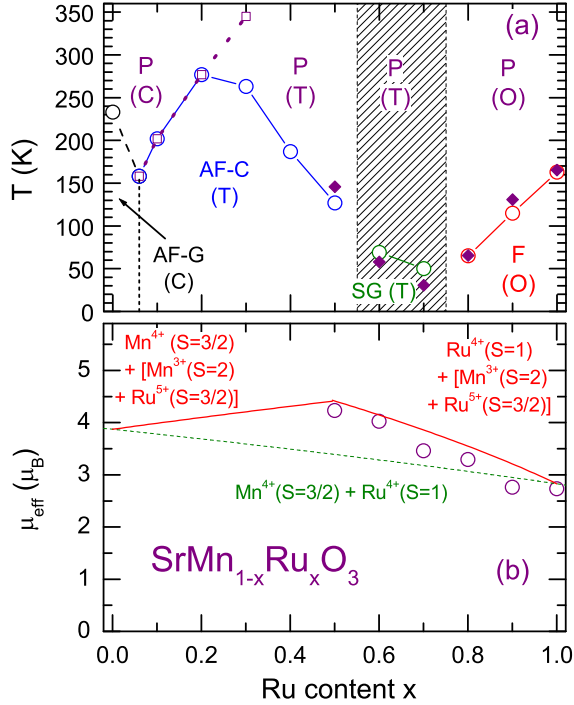} }
\caption{\label{phd} (Color online) (a) Phase diagram for  SrMn$_{1-x}$Ru$_{x}$O$_3$. Labels denote magnetic states [P, AF-G, AF-C, SG, and F are
paramagnetic, antiferromagnetic (type G), antiferromagnetic (type C), spin glass, and ferromagnetic, respectively]. Labels in parentheses denote
crystal symmetry (C, T, O are cubic, tetragonal, and orthorhombic, respectively). Open circles are magnetic transitions. Full circles are structural
cubic to tetragonal transitions. Full diamonds are paramagnetic Curie-Weiss temperatures. The hatched area is an approximate boundary of magnetically
frustrated region between the AF-C and F states. (b) The effective paramagnetic moment $\mu_{\rm eff}$ for SrMn$_{1-x}$Ru$_{x}$O$_3$. The lines
denote various models describing possible oxidation states of Ru and Mn ions (see:text)}
\end{figure}

The phase diagram in the low Ru substitution regime presented in Fig.~\ref{phd} (a), strikingly resembles the phase diagram for
$R_{y}$Sr$_{1-y}$MnO$_3$ ($R$=La, Pr).\cite{Chmaissem03} In both classes of materials, the two different substitutions in the parent SrMnO$_3$
compound, A-site and B-site, respectively change the band filling by generating exactly the same amount of Mn$^{3+}$ ions for the same substitution
level. The concentration of these ions is not sufficient to induce Mn$^{3+}$ - Mn$^{4+}$ DE interaction, but can induce the same tetragonal lattice
distortion, coupled with a C-type AF transition. The only magnetic ions present in La$_{y}$Sr$_{1-y}$MnO$_3$ are Mn$^{3+}$/Mn$^{4+}$ ions in the
ratio $y/(1-y)$. The $d$ shell electronic configuration of Ru$^{5+}$ ions in SrMn$_{1-x}$Ru$_{x}$O$_3$ is identical with that of Mn$^{4+}$ ($d^3$,
$t_{2g}^3$). Therefore, the ratio of $d^4$/$d^3$ ions in both classes of materials is identical, which leads to a very similar structural and
magnetic behavior. The paramagnetic Curie-Weiss temperature $\Theta$, also presented in Fig.~\ref{phd} (a), was calculated from the molar dc
susceptibility $\chi_m=M/H$ in the temperature range 350-400 K, which was fitted to the general Curie-Weiss formula:
\begin{equation}
 \chi_m = \chi_0 + (\mu_B N_A/3k_B)\mu_{\rm eff}^2/(T-\Theta),
\end{equation}
where $\chi_0$ is a temperature-independent background susceptibility, $N_A$ is the Avogadro number, $k_B$ is the Boltzmann constant, $\Theta$ is the
paramagnetic Curie-Weiss temperature, $\mu_{\rm eff}=g\sqrt{S(S+1)}$ is the effective paramagnetic moment, $g=2$ is the gyromagnetic ratio, $S$ is
the magnetic spin. The values of $\Theta$ pretty well coincide with the values of $T_C$ or $T_f$.

The effective paramagnetic moment $\mu_{\rm eff}$ determined from the magnetization using Eq.~(1) is presented in Fig.~\ref{phd} (b). We consider two
possible valence states of the Ru dopant, 4+ and 5+. The former case would lead to the following formula SrMn$^{4+}_{1-x}$Ru$^{4+}_{x}$O$_3$ and the
expected dependence $\mu_{\rm eff}=\sqrt{\mu^2_{\rm eff}{\rm (Mn)}+\mu^2_{\rm eff}{\rm (Ru)}}$, plotted as the dashed line in Fig.~\ref{phd} (b) is
far from the observed $\mu_{\rm eff}$ behavior. The latter case would give the following formulas SrMn$^{4+}_{1-2x}$Mn$^{3+}_{x}$Ru$^{5+}_{x}$O$_3$
and SrMn$^{3+}_{1-x}$Ru$^{5+}_{1-x}$Ru$^{4+}_{2x-1}$O$_3$ for $x\leqslant 0.5$ and $x\geqslant 0.5$, respectively. These formulae are plotted as
solid lines. In both cases we assume spin only moments. The latter model works well for $x\geqslant 0.5$. This is an additional evidence for the
presence of Ru$^{5+}$ ions in this material. However, significant deviations from any of the discussed models can be observed for  $x < 0.5$,
especially for pure SrMnO$_3$. The determined $\mu_{\rm eff}$ values are much lower in this region (not shown). $\Theta$ is also positive in the  $x
< 0.5$ solid solution range, which in turn is a sign of ferromagnetic interactions in the paramagnetic state, even if these compositions exhibit an
AF order at low temperatures. It is possible that in this doping regime the fitting temperature range is very close to the magnetic and structural
transition temperatures and the Curie-Weiss approximation in Eq.~(1) is not fully valid. Another explanation for reduced $\mu_{\rm eff}$ of SrMnO$_3$
and lightly substituted compositions may relate to unusual bond distances observed for these compounds that would require further study.

\subsection{Resistivity}
\begin{figure}
 \resizebox{8.5cm}{!} {\includegraphics{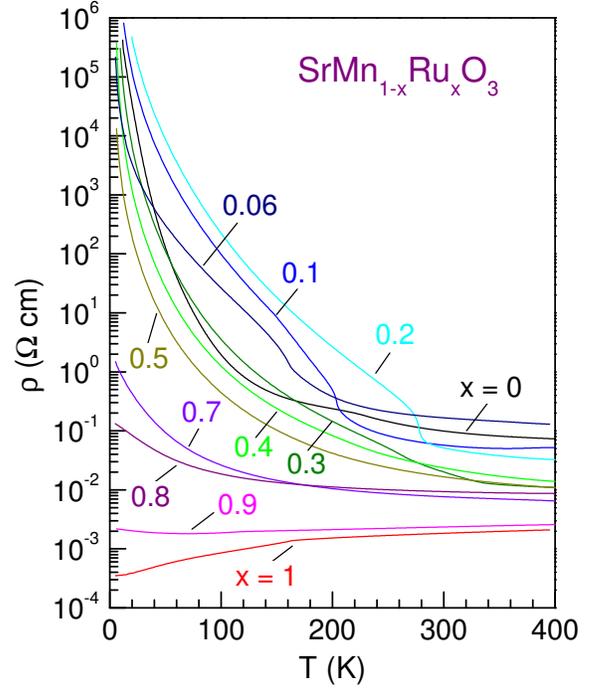} }
 \caption{\label{res} (Color online) Temperature dependence of resistivity  for  SrMn$_{1-x}$Ru$_{x}$O$_3$ samples.}
\end{figure}

\begin{figure}
 \resizebox{8.5cm}{!} {\includegraphics{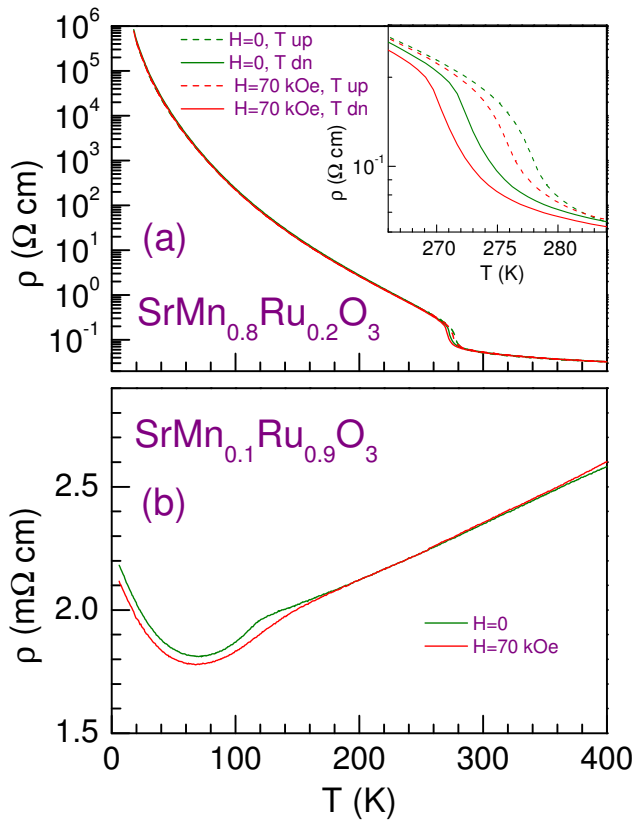} }
 \caption{\label{mr} (Color online) Magnetoresistance for selected SrMn$_{1-x}$Ru$_{x}$O$_3$ samples. }
\end{figure}

The temperature dependence of resistivity  $\rho(T)$ for  SrMn$_{1-x}$Ru$_{x}$O$_3$ samples is presented in Fig.~\ref{res}.
The resistivity demonstrates an insulating character for SrMnO$_3$ and overly decreases with increasing Ru substitution due to itinerant character of
the Ru electrons. It becomes metallic for $x \geqslant 0.9$ with a metal to insulator transition at around 70 K for $x=0.9$. The difference between
the characters of resistivity for highly Ru substituted polycrystalline samples and single crystals can be explained by the granular nature of the
polycrystalline samples. It has been demonstrated, e.g, for polycrystalline colossal magnetoresistance manganites that the presence of grain
boundaries can affect the magnitude of resistivity as well as the low-temperature magnetoresistance, without affecting their magnetic properties. For
low Ru contents ($x \leqslant 0.2$) a significant increase of resistivity is observed below the coupled AF-structural transitions. These transitions
correspond to a jump in $\rho(T)$ again similar to rare earth substituted SrMnO$_3$.\cite{Chmaissem03} This behavior corresponds to a first-order
phase transition with a hysteretic behavior of $\rho$ as illustrated in Fig.~\ref{mr}(a). The transition can be shifted to lower temperatures by
applying a magnetic field. A slight magnetoresistance can be observed for higher Ru contents below the Curie temperature. Generally, this effect is
rather small although enhanced with respect to pure SrRuO$_3$\cite{Banerjee01}. For $x = 0.3$, where the structural and magnetic transitions are
decoupled, a subtle anomaly in resistivity can be observed at a temperature related to the structural transition along with a smooth resistivity
behavior at the AF transition.

\subsection{Thermoelectric properties}

\begin{figure}
 \resizebox{8.5cm}{!} {\includegraphics{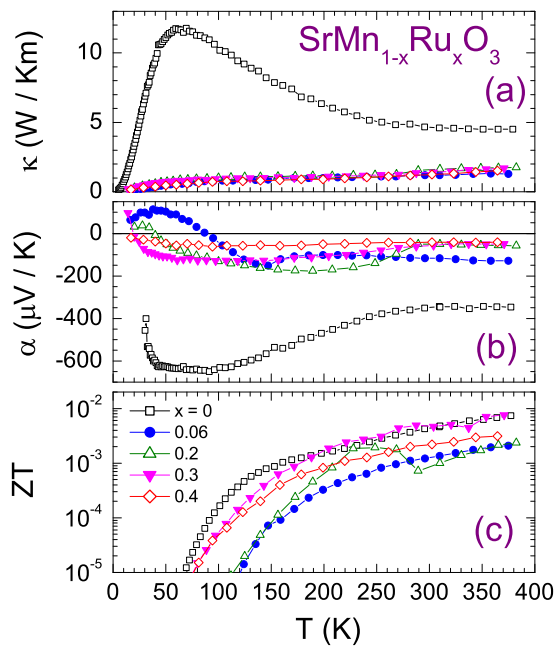} }
\caption{\label{tto} (Color online) (a) Thermal conductivity, (b) Seebeck coefficient, (c) Thermoelectric figure of merit $ZT=\alpha^2
T/(\kappa\rho)$ for selected SrMn$_{1-x}$Ru$_{x}$O$_3$ samples. }
\end{figure}
For higher Ru contents it has been demonstrated\cite{Klein06} that the Seebeck coefficient is positive and varies slightly from $+34 \mu$V/K for
SrRuO$_3$ to $+28 \mu$V/K for SrMn$_{0.1}$Ru$_{0.9}$O$_3$ at room temperature. This small change reflects the introduction of Ru$^{5+}$ ions into the
Ru$^{4+}$ matrix. In Fig.~\ref{tto}, we present thermal conductivity $\kappa$, Seebeck coefficient $\alpha$, and the thermoelectric figure of merit
$ZT=\alpha^2 T/(\kappa\rho)$ for selected SrMn$_{1-x}$Ru$_{x}$O$_3$ samples. For SrMnO$_3$, $\alpha$ is large and negative ($-350 \mu$V/K at RT). A
low Ru substitution induces a drastic change of $\alpha$ to values of -50 to -60 $\mu$V/K at RT. This negative effect on the thermoelectric
properties is compensated by a significant decrease in $\rho$ and $\kappa$, which, e.g., gives a similar values of $ZT$ at and above room temperature
for $x=0.3$ as for pure SrMnO$_3$. We observe a crossover of $\alpha$ from negative to positive values at low temperatures for a low Ru substitution.
This crossover shifts to lower temperatures with the Ru substitution. A similar effect has been seen in SrMnO$_3$ with a different $B$-site
substitution (Mo) as well as with an $A$-site substitution (Pr).\cite{Maignan02b}

\section{SUMMARY}
In summary, we have studied the phase diagram of polycrystalline perovskite SrMn$_{1-x}$Ru$_{x}$O$_3$ ($0 \leqslant x \leqslant 1$) system. In the
low Ru$^{5+}$ substitution regime ($x \leqslant 0.3$), the structural, magnetic, and transport behavior strikingly resemble those for the SrMnO$_3$
compound with an A-site heterovalent substitution  $R_{y}$Sr$_{1-y}$MnO$_3$ ($R$=La, Pr). In both cases, a tetragonal lattice distortion, for some
compositions coupled to a C-type AF transition is observed. This similarity is driven by the same ratio of $d^4$/$d^3$ ions in both classes of
materials for equivalent substitution level. In the moderate Ru$^{5+}$ substitution regime ($x \sim 0.65$) a boundary between the AF-C and F orders
in polycrystalline  SrMn$_{1-x}$Ru$_{x}$O$_3$  is broadened with respect to a sharp quantum critical point previously observed in single crystals due
to magnetic disorder, which leads to a spin glass behavior. The observation of a spin glass behavior suggests that the AF-C and F states are
separated by a first-order transition in the clean limit and they can coexist in the presence of quenched disorder.\cite{Burgy01}

\section*{ACKNOWLEDGMENTS}
Work at NIU was supported by the NSF (DMR-0706610). Work at Argonne National Laboratory was supported by the U. S. Department of Energy under
contract DE-AC02-06CH11357.

\end{document}